          [2001/04/25 1.1 (PWD)]
\documentclass[a4paper,twocolumn]{esapub} 
\usepackage{graphicx,natbib}

\title{The Nature and Excitation Mechanisms of Acoustic 
Oscillations in Solar and Stellar Coronal Loops}
\author{D. Tsiklauri$^1$, V.M. Nakariakov$^2$, 
 T.D. Arber$^2$,  M.J. Aschwanden$^3$}
\affil{$^1$Institute for Materials Research, School of CSE,
University of Salford, Gt Manchester, M5 4WT, UK\\
$^2$Physics Department, University of Warwick, Coventry, CV4 7AL,
England, UK \\
$^3$LMSAL, Dept. L9-41, Bldg.252
3251 Hanover Street, Palo Alto, CA 94304, USA}

\begin{document}


\maketitle

\begin{abstract}
In the recent work of Nakariakov et al. (2004),
it has been shown that 
the time dependences of density and velocity
in a flaring loop contain pronounced quasi-harmonic 
oscillations
associated with the 2nd harmonic of a standing slow magnetoacoustic wave.
That model used a symmetric
heating function (heat deposition was strictly at the apex).
This left outstanding questions: A) is the generation
of the 2nd harmonic a consequence of the fact that the heating function was
symmetric? B) Would the generation of these oscillations occur
if we break symmetry? C) What is the spectrum of these 
oscillations? Is it consistent with a 2nd spatial harmonic?
The present work (and partly Tsiklauri et al. (2004b))
attempts to answer these important outstanding questions. 
Namely, we investigate the physical nature of these oscillations
in greater detail: we study  their spectrum (using periodogram technique)
and how heat positioning affects the mode excitation.
We found that excitation of such oscillations is
practically independent of location of the
heat deposition in the loop. Because of the change of the
background temperature and density, the phase shift between the
density and velocity perturbations is not exactly a
quarter of the period, it varies along the loop and
is time dependent, especially in the case of one footpoint (asymmetric)
heating. We also were able to model successfully SUMER oscillations
observed in hot coronal loops.
\end{abstract}

\section{Introduction}
Magnetohydrodynamic (MHD) coronal seismology is one of the main
reasons for 
studying waves  in the solar corona. Such
studies also are important in connection
with coronal heating and solar wind acceleration problems. 
Observational evidence of coronal waves and oscillations in EUV 
are numerous (e.g. \citep{o99,ow02,wang02,wang03}). 
Radio
band observations also demonstrate various kinds 
of oscillations (e.g., the
quasi-periodic pulsations, or QPP, see \citet{a87} for a
review), usually with periods from a few seconds to tens 
of seconds.  Decimetre and microwave observations  show much
longer periodicities, often in association with a flare. For
example, \citet{wx00} observed QPP with the periods of about
50~s at 1.42 and 2~GHz (in association with an M4.4 X-ray flare).
Similar periodicities have been observed in the X-ray band  (e.g.
\citep{m97,terekhov02}) and in 
the white-light emission associated with the stellar
flaring loops \citep{mathio}.
A possible interpretation of these medium period QPPs may be
in terms of kink or torsional modes \citep{zs89}. 

In our previous, preliminary  study 
 \citep{nta04}, we outlined an alternative,  simpler, thus more
attractive mechanism for the generation
of long-period QPPs. That model used a symmetric
heating function (heat deposition was strictly at the apex).
This left the outstanding questions: A) is the generation
of the 2nd harmonic a consequence of the fact that the heating function was
symmetric? B) Would the generation of these oscillations occur
if we break symmetry? C) What is the spectrum of these 
oscillations? Is it consistent with a 2nd spatial harmonic?
The present work (and partly \citet{tb04}) attempts to answer these important 
outstanding questions. 

We also were able to model successfully SUMER oscillations
observed in hot coronal loops.

The paper is organised as follows: in sect. 2 we
present the numerical results, with subsection 2.1
dedicated to the case of apex (symmetric) heating
which completes the work started in \cite{nta04},
and subsection 2.2 summarising our findings in the 
case of single footpoint (asymmetric) heating.
In sect.3 we present some preliminary results of modelling of SUMER oscillations
observed in hot coronal loops.
We close with conclusions in sect. 4.

\section{Numerical Results}

The model that we use to describe plasma dynamics in a coronal loop
is outlined in \citet{nta04,ta04}. Here we just add that,
when numerically solving
 the 1D radiative hydrodynamic equations (infinite magnetic
field approximation), and using a 1D version of the Lagrangian Re-map
code (Arber et al. 2001) with the radiative loss limiters, 
the radiative loss function was specified as in \citet{ta04}
which essentially is the \citet{rtv78} law extended to a wider temperature range
\citep{pe82,p82}.

We have used the same heating function as in \citet{ta04}.
The choice  of the temporal
part of the heating function is such that at all times
there is a small background heating present (either at footpoints or
the loop apex) which ensures that in the absence of flare heating
(when $\alpha$, which determines the flare heating amplitude, is zero)
the average loop temperature
stays at 1 MK.
For easy comparison between the apex and footpoint heating
cases we fix $Q_p$, flare heating amplitude, at a  different 
value in each case (which ensures that with the flare heating on
when $\alpha=1$
the average loop temperature
peaks at about the observed value of 30 MK in { both cases}).

In the numerical runs in sect.2,  $1/(2 \sigma_s^2)$
was fixed at 0.01 Mm$^{-2}$, which gives a heat deposition length scale, 
$\sigma_s=7$ Mm. This is a typical value determined from
the observations  \citep{abk02}. 
The flare peak time was fixed in all numerical
simulations at 2200 s. 
The duration of the flare, $\sigma_t$, was fixed at 333 s.
The time step of data visualisation
was chosen to be 0.5 s. The CFL limited 
time-step used in the simulations was 0.034 s. 

\subsection{Case of Apex (Symmetric) Heating}

In this subsection we
complete the analysis started in \citet{nta04}, namely for the same 
numerical run we study the spectrum of oscillations
at different spatial points.

\begin{figure}[]
\resizebox{\hsize}{!}{\includegraphics{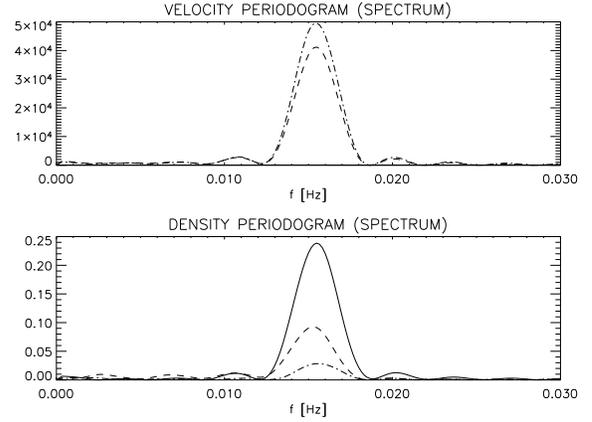}} 
\caption{Case of apex (symmetric) heating: Periodogram
(spectrum) of the velocity and density oscillatory
component times series outputted in the following three points: 
loop apex
 (solid curve), 
$1/4$ (dash-dotted curve) and $1/6$ (dashed curve)
of the effective loop length (48 Mm), i.e. at $s=0,-12,-16$ Mm.} 
\end{figure}

As was pointed out in 
\citet{nta04},
the most interesting fact is that we see clear quasi-periodic
oscillations, especially in the second stage (peak of the flare)
for the time interval $t=2500-2800$ s (cf. Fig.~1 in \citet{nta04}).
Such oscillations are frequently seen 
during the solar flares observed in X-rays, 8-20 keV, 
(e.g. \citet{terekhov02}) as well as stellar flares observed
in white-light (e.g. \citet{mathio}).

Before discussing the
physical nature of these oscillations,
it is worth recalling for completeness the simple 1D analytic theory
of standing sound waves. For 1D, linearised, hydrodynamic 
equations with constant unperturbed (zero order) background variables,
the solutions for density, $\rho$, and velocity, $V_x$, can be
easily written as
\begin{equation}
V_x(s,t)=A \cos \left( {{n \pi C_s}\over{L}} t\right)
\sin \left( {{n \pi}\over{L}} s\right),
\end{equation}
\begin{equation}
\rho(s,t)=-{{A \rho_0}\over{C_s}} \sin \left( {{n \pi C_s}\over{L}} t\right)
\cos \left( {{n \pi}\over{L}} s\right),
\end{equation}
where $C_s$ is the speed of sound, $A$ is wave amplitude, $L$ is
loop length, $n=1,2,3,...$ is the harmonic number, and $s$ is the distance
along the loop.
Note the (relative) phase shift between $V_x$ and $\rho$
is $\Delta P/ P=-(\pi/2)/(2 \pi)=-1/4$, where $P$ is the standing wave
period, while this ratio is zero for a propagating wave.
  Also, Eqs.(3) and (4) from \citet{nta04} are missing
a factor of $2$, while our Eqs.(1) and (2) correct this 
previous omission.

\begin{figure}[]
\resizebox{\hsize}{!}{\includegraphics{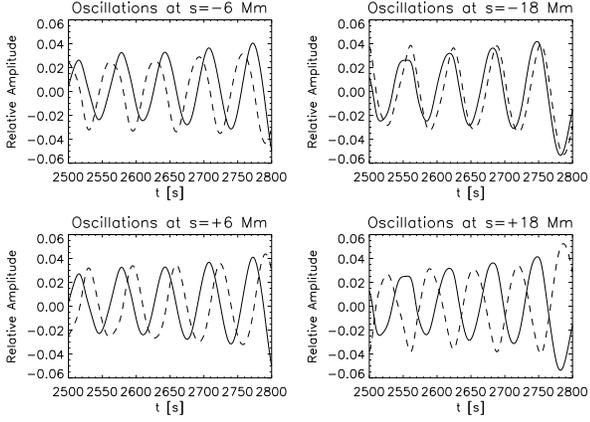}} 
\caption{Case of apex (symmetric) heating:
oscillatory components of time series outputted at 
$\pm 6$ Mm and $\pm 18$ Mm in   
the time interval of 2500-2800 s.
The solid curve shows
plasma number density in units of $10^{11}$ cm$^{-3}$. 
The dashed curve shows velocity  normalised to 400 km s$^{-1}$.} 
\end{figure}

 In Fig.~1 we present  a periodogram
(which here we use interchangeably with the (power) spectrum,
although strictly speaking the power spectrum is a theoretical
quantity defined as an integral over continuous time, and of which
the periodogram is simply an estimate based on a finite amount of discrete
data (cf. \citet{sca82} and his Eq.(10) in particular)) 
of the velocity and density time series outputted at the three points: 
loop apex, 
$1/4$ and $1/6$ of the effective loop length (48 Mm), 
i.e. at $s=0,-12,-16$ Mm.
The first two points are chosen to test whether
the simple analytic solution for 1D standing sound waves (see below)
is relevant in this case. The third point ($1/6$) was chosen
arbitrarily (any spatial point along the loop where density and
velocity of the standing waves does not have a node would be
equally acceptable).
As expected for a 2nd spatial
harmonic of a standing sound wave in the velocity periodogram
there are two clearly defined peaks and the largest peak
corresponds to $1/4$ of the effective loop length, while
the smaller peak corresponds to  $1/6$. Note that at the
loop apex the periodogram gives $0$ (solid line is too close to zero
to be seen in the plot). The density periodogram shows the
opposite behaviour to that of the velocity with the largest peak
corresponding to the loop apex, while $1/6$ of the effective loop length
corresponds to a smaller peak, and $1/4$ is close to zero. 
The locations of the peaks are at about 0.0155 Hz i.e. the period of
the oscillation is 64 s. The period of a 2nd spatial
harmonic of a standing sound wave should be 
\begin{equation}
P=L/C_s=L/(1.52 \times 10^5 \sqrt{T}),
\end{equation}
 where
 $T$ is plasma temperature measured in MK, while $L$ is in meters.
If we substitute an effective loop length $L=48$ Mm (see Fig.~2 in 
\citep{nta04}) and
an average temperature of 25 MK (see top panel in Fig.~1 in 
\citet{nta04} in the 
range of 2500-2800 s -- the quasi periodic oscillations time interval we study)
we obtain 63 s, which is  close to the result of our numerical
simulation. Such a close coincidence is surprising
bearing in mind that the theory does not take into
account variation of background density and velocity over time, while
we see from Fig.~1 in \citet{nta04} that even within a short interval
of a flare, i.e. 2500-2800 s, all physical quantities
vary  significantly with time.
To close our investigation of the physical nature of
the oscillations we study the phase shift between 
the velocity and density oscillations and
compare our simulation results with analytic theory.
In Fig.~2 we plot time series, outputted at 
$\pm 6$ and $\pm 18$ Mm,  
of the plasma number density in units of $10^{11}$ cm$^{-3}$
and velocity, normalised to 400 km s$^{-1}$. 
These points were selected so that one
symmetric (with respect to the apex) pair ($\pm 6$ Mm)
is close to the apex, while another pair
($\pm 18$ Mm) is closer to the footpoints.
 We choose these pairs because we wanted
to compare how the phase shift is affected by spatial location.
One would expect stronger upflows close to footpoints
(due to chromospheric evaporation), which in turn alters the phase
shift. Note that phase shift between the density and velocity
is different (see below) for standing and propagating (with flows)
acoustic waves).
We gather  from the graph that: (A) clear quasi-periodic
oscillations are present; (B) they are shifted with respect 
to each other in time;
(C) near the apex ($\pm 6$ Mm)
the phase shift is close to that
predicted by 1D analytic theory;
(D) close to the footpoints ($\pm 18$ Mm)
the phase shift is somewhat different from the one
predicted by 1D analytic theory.
In the last case the discrepancy can be attributed
to the presence of flows near the footpoints.
The main reason for the overall deviation is due to the
fact that analytic
theory does not take into
account variations of background density and velocity in time
and that density gradients in the
transition region are not providing perfect
reflecting boundary conditions for the
formation of standing sound waves.

Another interesting result is that even with the wide variation
of the parameter space of the flare, its duration, peak average
temperature, etc., we always obtained a
dominant 2nd spatial harmonic of a standing sound wave with 
some small admixture of 4th and sometimes 6th harmonics.
Our initial guess was that this is due to the symmetric excitation
of these oscillations (recall that we use apex heat deposition).
In order to investigate the issue of excitation further
we decided to break the symmetry and put the heating source at one footpoint,
hoping to see excitation of odd harmonics 1st, 3rd, etc.

\subsection{Case of Single Footpoint (Asymmetric) Heating}

For single footpoint heating we fix $s_0= 30$ Mm in Eq.(1) 
in \citet{nta04}, i.e.
(spatial) peak of the heating is chosen to be at the bottom 
of the transition region (top of chromosphere).
Initially we run a code without flare heating,
i.e. we put $\alpha=0$ (in this manner we turn off flare 
heating). $E_0=0.02$ erg cm$^{-3}$ s$^{-1}$ 
was chosen such that in the steady (non-flaring) case
the  average loop temperature stays at
about 1 MK.
Then, we run the code with flare heating,
i.e. we put $\alpha=1$, and fix $Q_p$ at $1 \times 10^4$,
so that it yields a peak average temperature of about 30 MK.
The results are presented in Fig.~3. 
During the
flare the apex temperature peaks at 38.38 MK while 
the number density at the apex peaks at $3.11 \times 10^{11}$ cm$^{-3}$.
In this case, as opposed to the case of symmetric (apex)
heating, the velocity dynamics is quite different.
Since the symmetry of heating is broken there is a non-zero
net flow through the apex at all times.
However, as in the symmetric heating case, 
we again see quasi-periodic
oscillations superimposed on the dynamics of all physical
quantities (cf time interval of 
$t=2400-2700$ in Fig.~3). 

In Fig.~4 we present time-distance
plots of
velocity  and density for the time 
interval 2400-2700s, where the quasi-periodic
oscillations are most clearly seen.
Here we again subtracted the slowly varying background (with respect to
oscillation period).
The picture is quite different from the case of apex (symmetric)
heating (compare it with Fig.~2 in \citep{nta04}). 
This is because now the
node in the velocity (at the apex) 
moves back and forth  periodically along the apex,
and at later times ($t>2550$ s) stronger flows
are now present.
However, the physical nature of the oscillations remains 
mainly the same. i.e. a 2nd spatial harmonic of a sound wave, 
but now with an oscillating node at the apex.

\begin{figure}[]
\resizebox{\hsize}{!}{\includegraphics{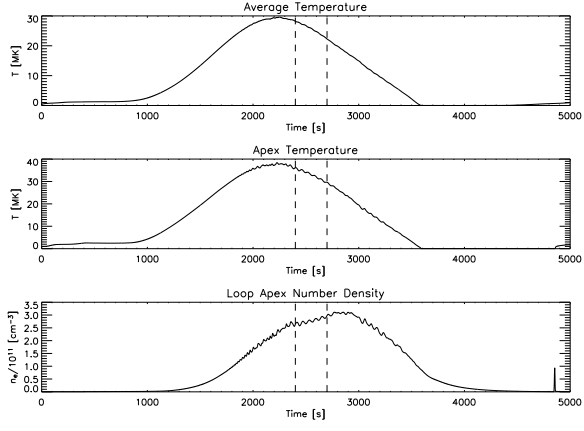}} 
\caption{Case of single footpoint (asymmetric) heating: 
Average temperature, temperature at apex, and 
 number density at the apex as
 functions of time.}
\end{figure}

To investigate this further we plot in Fig.~5 a periodogram
(spectrum) of the velocity and density oscillatory
component time series outputted at the following three points: 
loop apex, $1/6$ and $1/4$ of the effective loop length.
We gather from the graph that the  periodogram
(spectrum) is more complex than in the case of apex (symmetric)
heating. In the velocity periodogram
at the apex there is a peak with a frequency 
higher than that of 2nd spatial harmonic of
a standing sound wave. This is the frequency with
which the node of the velocity oscillates (see discussion in the 
previous paragraph). It has nothing to do with the standing
mode, but is dictated by the excitation conditions of the loop which
acts as a dynamic resonator. 
Let us analyse now how this periodogram compares
with 1D analytic theory. The peak in the periodogram
corresponding to $1/6$ of the effective loop length
(dashed line) corresponds to a frequency of about 0.017 Hz, i.e.
the period of oscillation is 59 s.
Again, the period of a 2nd spatial
harmonic of a standing sound wave should be 
calculated from Eq.~(3).
If we substitute the effective loop length $L=48$ Mm (see Fig.~4) and
an average temperature of 26 MK (see top panel in Fig.~3 in the 
range of 2400-2700 s)
we obtain 62 s, which is close to the result of our numerical
simulation (59 s).

\begin{figure}[]
\resizebox{\hsize}{!}{\includegraphics{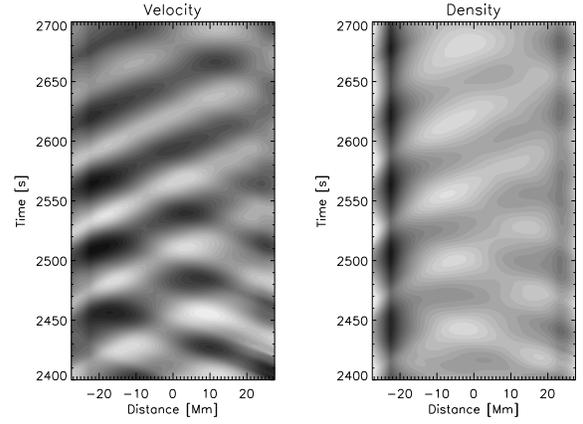}} 
\caption{Time-distance plots of the velocity and 
density oscillatory components in the
time interval of 2400-2700s for the case of single footpoint (asymmetric) 
heating.} 
\end{figure}

Next, we studied the phase shift between velocity and density
oscillations, and
compare our simulation results with the 1D analytic theory.
In Fig.~6 we produce a plot similar to Fig.~2,
but for the case of asymmetric heating.
The deviation, which is greater than in the case of apex (symmetric)
heating, can again be attributed to the over-simplification of the
1D analytic theory, which does not take into account time variation
of the background physical quantities and 
imperfection of the reflecting boundary conditions
(see above).
More importantly, in the asymmetric case strong
flows are present throughout the flare simulation time.
Thus, if linear time dependence is assumed,
which is relevant within the short interval 2400-2700 s
of the flare, then Eqs.(1)-(2) would
be modified such that phase shifts would vary 
secularly in time. This is similar to that seen in Fig.~6.

\begin{figure}[]
\resizebox{\hsize}{!}{\includegraphics{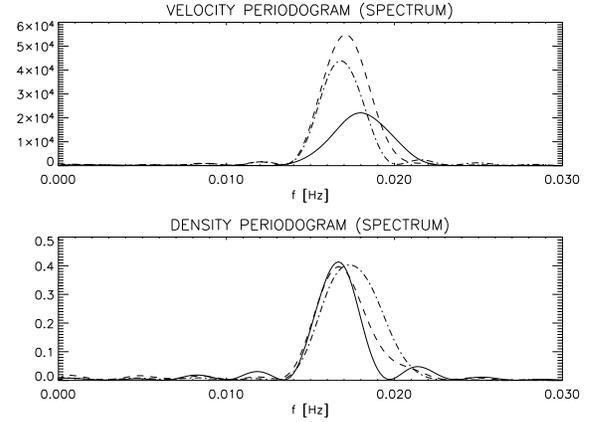}} 
\caption{As is Fig.~1 but for the
case of single footpoint (asymmetric) heating.
Time interval here is 2400-2700s.} 
\end{figure}

Yet another interesting observation comes from the
following argument: in a steady 1D case analytic theory
predicts that the phase shift between the density
and velocity should be (A) zero for for a { propagating} 
acoustic wave and (B) quarter of a period for the
{ standing} acoustic wave.
Since in the asymmetric case strong flows are present,
we see less phase shift between the velocity and
density in Fig.~6 as one would expect.

\begin{figure}[]
\resizebox{\hsize}{!}{\includegraphics{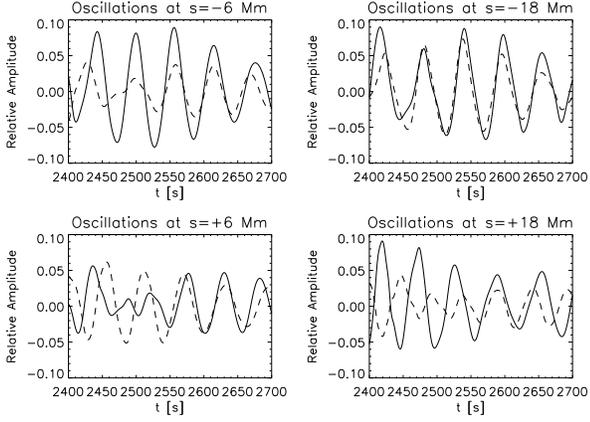}} 
\caption{As in Fig.~2, but for the case of single footpoint (asymmetric) 
heating. Time interval here is 2400-2700s.} 
\end{figure}

Thus, the results of the present study (and partly \citet{tb04})
provide further, and { more definitive} proof than in \citet{nta04}
that these oscillations are indeed
the 2nd spatial harmonic of a standing sound wave. However, the
present work also reveals that in the case
of single footpoint (asymmetric) heating the physical nature
of the oscillations is more complex, as the node in the velocity
oscillates along the apex and net flows are present.

\begin{figure}[]
\resizebox{\hsize}{!}{\includegraphics{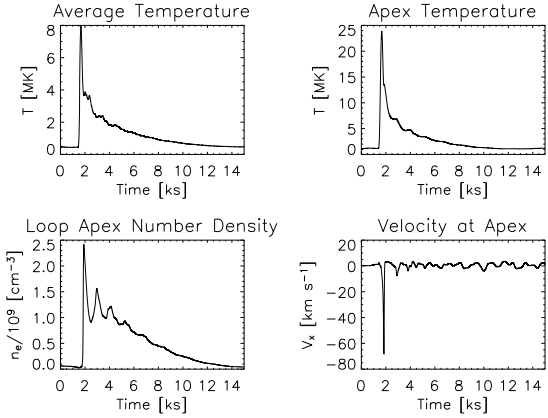}} 
\caption{Case of apex (symmetric) heating: 
Average temperature, temperature at apex,  
 number density at the apex, and velocity at the apex as
 functions of time.} 
\end{figure}

\begin{figure}[]
\resizebox{\hsize}{!}{\includegraphics{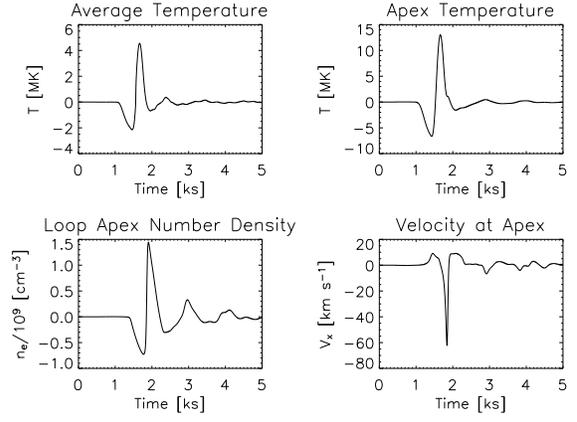}} 
\caption{As in Fig.~7, but with with subtracted
slowly varying background.} 
\end{figure}

\section{SUMER oscillations of hot coronal loops}

Recently hot ($T=6.3$ MK and $T=8$ MK) coronal loop oscillations were observed with SUMER
(cf. for review \citet{wang03}). These were interpreted as standing acoustic waves \citet{ow02}.
So far only highly simplified models (isothermal, no stratification, no Helium presence,
no external heating input, no transition region or chromosphere, etc.) were used to study these oscillations.
Excitation of the standing waves was done artificially perturbing velocity
of plasma, globally, along the loop and then their evolution (rapid decay)
 was modelled numerically with heat conduction being dominant factor \citet{ow02}.
Here we use much more sophisticated model, as described above, to self-consistently
excite these oscillations by adding external heat input.
In a similar manner as in previous sections we use apex (symmetric)
heating. Parameters of the heating function were chosen such  the average
temperature peaks at about $8$ MK. The time dynamics of various quantities are
shown in Fig.7. In Fig. 8 we plot the same quantities with subtracted
slowly varying background. This shows clear, periodic, decaying oscillations
are excited. Loop length was fixed at $L=350$ Mm.

From Fig.~8 it follows that the period of oscillations is
about 800 s $\approx 13$ min, which is similar to the
observed range of $17.6 \pm 5.4$ min SUMER oscillations
(based on 54 Doppler-shift cases \citep{wang03}).
Moreover, the same result is obtained from the simple 1D
analytic theory when  loop length $L=350$ Mm and temperature of
$8$ MK is substituted into Eq.(3).
Thus, our model seems to describe adequately essential observed
features of these oscillations.
Parametric study showed that in the case SUMER
oscillations heating function should have fairly short
temporal width (about 1 minute).
The main difference from previous models is that we 
self-consistently deal with the excitation problem.

Note also that in this simulation period of the oscillations
is increased. This can be attributed to two factors:
(A) decrease of backround tempreature (which would not be observed
by SUMER because it looks at lines with fixed temperatures ($T=6.3$ MK and $T=8$ MK)),
and (B) dissipative effects (mainly heat conduction).
It seems that the increase in period observed by SUMER should be attributed to the
latter effect (B).

\section{Conclusions}

Initially we used a 1D radiative hydrodynamics 
loop model which incorporates
the effects of gravitational
stratification,
heat conduction, radiative losses, added external 
heat input, presence of helium, hydrodynamic 
non-linearity, and bulk Braginskii viscosity
to simulate flares \citep{ta04}. 
As a byproduct of that study,
in practically all the numerical runs 
quasi-periodic oscillations in all physical quantities
were detected \citep{nta04}.
Such oscillations are frequently seen 
during the solar flares observed in X-rays, 8-20 keV 
(e.g. \citet{terekhov02}) as well as stellar flares observed
in white-light (e.g. \citet{mathio}).
Our present analysis (and partly \citet{tb04}) shows that quasi-periodic
oscillations seen in our numerical simulations
bear many similar features compared to observational datasets.
In this work we tried to answer 
important {\it outstanding questions} (cf. Introduction section)
that arose from the previous analysis \citep{nta04}.

In summary the present study (and \citet{nta04,tb04})  established
 the following features:
\begin{itemize}
\item We show that the time dependences of density and temperature
 in a flaring loops contain well-pronounced quasi-harmonic oscillations
 associated with standing slow magnetoacoustic modes of the loop.
\item
For { a wide range of physical parameters}, 
the dominant mode is the second spatial harmonic, with a velocity
 oscillation node and the density oscillation maximum at the loop apex.
{ This result is practically independent of the positioning of the
 heat deposition in the loop}. 
\item 
 Because of the change of the
 background temperature and density, 
 and the fact that 
density gradients in the
transition region are not providing perfect
reflecting boundary conditions for the
formation of standing sound waves,
 the phase shift between the
 density and velocity perturbations is not exactly equal to a
 quarter of a period.
\item 
 We conclude that the oscillations in the white light,
 radio and X-ray light curves observed during
 solar and stellar flares may be produced by slow standing modes,
with the period determined by the loop temperature and length. 
\item
For a typical
 solar flaring loop the period of oscillations is shown to be about
 a few minutes, while amplitudes are typically of a few percent.
 \item
Our model seems to describe adequately essential observed
features of SUMER oscillations.
Parametric study showed that in this case
heating function should have fairly short
temporal width (about 1 minute).
\end{itemize}

The novelty of this study is that by studying the spectrum 
and phase shift of these oscillations we provide more definite proof
that these oscillations are indeed the 2nd harmonic of a standing sound wave,
and that the single footpoint (asymmetric) heat positioning
still produces 2nd spatial harmonics, although it is more complex than the
apex (symmetric) heating as the node in the velocity
oscillates along the apex and net flows are also present. 

\section*{Acknowledgments}
DT kindly acknowledges support from Nuffield Foundation 
through an award to newly appointed lecturers in science,
engineering and mathematics (NUF-NAL 04).

\end{document}